\documentclass[prl,preprint]{revtex4}
\usepackage{color}
\usepackage{amsfonts}

\begin{document}

\title{Quantum Confinement of Bloch Waves\\ and \\Related Problems}

\author{Shang Yuan Ren}
\address{School of Physics, Peking University, Beijing 100871, P.R. China}

\email{shangyuanren21@gmail.com}

\maketitle

{\center{\center Abstract}}
~\newline
The quantum confinement of Bloch waves is fundamentally different from the well-known quantum confinement of plane waves.
Unlike that obtained in the latter are all stationary states only; in the former, there is always a new type of states
--- the boundary dependent states.
\par
This distinction leads to interesting physics in low-dimensional systems.

\vskip 3in
~~\\
{\bf APS Physics Subject Headings}:
\\
{\bf DISCIPLINES}: General Physics. \\
{\bf Research Areas}: Quantum theory.\\
                {\bf Concepts}: Non-relativistic wave equation.\\
{\bf Theoretical Techniques}:
Mathematical physics methods.

\vskip .02in
~~\\
{\bf DISCIPLINES}: Condensed Matter \& Materials Physics.\\
{\bf Research Areas}: Electronic structure: Surface states, Edge states.\\
{\bf Physical Systems}: Quantum wells, Thin films, Quantum wires, Quantum dots.\\
{\bf Theoretical Techniques}:
Theory of Periodic Differential Equations.

\newpage
\section{1. Quantum confinement is a fundamental problem in quantum mechanics}
Quantum confinement is one of the fundamental problems
in quantum mechanics. A problem discussed in many quantum mechanics textbooks  ---
the square potential well problem\cite{lan} --- is essentially a problem of
the quantum confinement of well-known plane waves. It has long been a well-understood
problem: when the electron is completely confined in a square
potential well of a specific width, all possible electronic states are stationary
wave states. The energy of each state (from the bottom of the potential well) can
only take positive discrete values, and they all increase as the well-width decreases.
Concepts and basic understandings obtained from the problem are often applied in
related quantum mechanics problems.

\section{2. Bloch wave is a more general type of wave}

Bloch's theory \cite{blh} is the very basis of modern solid state physics.
The Bloch wave is the most fundamental and essential concept in modern solid state
physics. It is a more general type of wave than the well-known plane
wave: a Bloch wave becomes a plane wave when the potential period(s) involved goes zero.

\par
The theory of electronic states based on Bloch's theory is essentially a theory
of electronic states in infinite crystal: only the arrangement of all atoms in the infinite
crystal has periodic translational invariance on which Bloch's theorem is based. Because
the size of any real crystal is finite, this conventional theory has some fundamental
difficulties. The difficulties become more significant nowadays since one might often face
crystals of a much smaller size than those in the past, such as submicron and nanoscales.

\par
Theoretically, a problem on electronic states in a low-dimensional system or a finite crystal
can also be considered a problem of quantum confinement of Bloch waves in a specific space,
with a specific boundary and a specific size.
To Bloch wave such a more general type of wave, the solid state physics community knows
mathematically quite limited besides some most basic understandings.
Therefore, for Bloch waves' quantum confinement, one of the most common practices
is to directly adopt the well-known treatment of plane waves' quantum confinement,
only with a specific effective mass determined by the particular crystal as the electron
mass in the treatment.
This practice is the frequently used effective mass approximation method.
Strictly speaking, such a direct adoption is not theoretically justified.
Nevertheless, due to the lack of clear understanding of the Bloch wave,
such practices are common.
The results obtained are sometimes far from the experimental results
or numerical methods and might even be completely
contradictory. The reason is simply that, in general, a Bloch wave is not a plane wave.

\section{3. Complete quantum confinement of Bloch wave may produce two different types of electronic states}

Based on the mathematical theory of periodic differential equations \cite{eas}, the author
developed an analytical theory on Bloch waves' complete quantum confinement \cite{syr1,stmp1,stmp2}.
\par
The Bloch waves' complete quantum confinement in one-dimensional space can be
rigorously solved analytically \cite{syr1,stmp1,stmp2}.
The complete quantum confinement of Bloch waves in a multi-dimensional space
is, in general, more difficult.
Nevertheless, many essential conclusions can be derived based on mathematical
theorems with help from physical intuitions \cite{stmp1,stmp2}.
\par
The most fundamental understanding obtained is that in some simple but essential cases,
the Bloch waves' complete quantum confinement once (in one direction) may produce
two different types of electronic states.
One is the states that depend on the confinement boundary,
and the other is the states that depend on the confinement size.
\par
The states that depend on the size are the stationary states of the Bloch wave.
Their numbers, properties, and energies depend on the
confinement size but not on boundary position.
For each band of the Bloch waves, there {\it always exists one}
state or sub-band of a new type, whose property and energy
depend on the confinement boundary position but not size and might be
in forbidden bandgap where the Bloch wave cannot exist.
\par
Such an understanding of the complete quantum confinement of Bloch waves in general
one-dimensional cases as well as in some simple and essential multi-dimensional
cases can be summarized in a mathematical form as\cite{stmp2}
\begin{equation}
N  = {\bf{\color{red} 1}} + \textit{\color{blue}(N~-~1)}.
\label{eqn:EQ1}
\end{equation}
\par
Eq. (\ref{eqn:EQ1}) indicates that in a truncated periodic structure
contains ${N}$ periods (left), for {\it each permitted band}
of the Bloch wave, there are two different types of states or sub-bands
(right) in the truncated structure. There is ${\bf{\color{red} 1}}$
state or sub-band whose property and energy depend on the boundary,
and there are $\textit{{\color{blue}(N~-~1)}}$ stationary Bloch
wave states or sub-bands whose  properties and energy depend on the
structure size $N$.
\par
The boldfaced ${\bf{\color{red} 1}}$ state (or sub-band) in Eq. (1) indicates
the very distinct point of Bloch waves' quantum confinement.
It is precisely due to the always existence of the boundary-dependent
${\bf{\color{red} 1}}$ state or sub-band that makes the Bloch waves'
quantum confinement significantly different from both
the plane waves' quantum confinement --- all states are
size dependent stationary states --- and the results of the periodic boundary
conditions conventionally used in solid state physics --- all states are Bloch
waves, no electronic states in forbidden band-gap.
The well-known surface states are among the states dependent on the boundary.
Therefore, the boundary-dependent states' existence is a fundamental
distinction of the Bloch waves' quantum confinement.
\par
In multi-dimensional space confinement, the boldfaced ${\bf{\color{red} 1}}$ state
or sub-band depending on the boundary always exists.
Instead, the other electronic states are less understood except for some simple
and essential cases.
We have understood multi-dimensional Bloch stationary states neither mathematically
nor physically enough for further sound general scientific reasonings.
\par
Bloch wave is more general than the plane wave. The new theory of Bloch waves'
complete quantum confinement is more general than the corresponding theory on
plane waves --- the infinite-deep square potential well problem in quantum mechanics.
An analytically solvable problem in quantum mechanics can often be of scientific
value. The complete quantum confinement of one-dimensional Bloch waves can be
rigorously solved analytically, adding an interesting example.

\section{4. Existence and property of boundary dependent states make low-dimensional systems more abundant in physics}

\par
Since the last half-century, research on low-dimensional systems (such as quantum wells, quantum wires, quantum dots)
in the sub-micron and nanometer scales has developed rapidly.
Experiments showed that in these low-dimensional systems, the semiconductor crystal's
physical properties could change significantly as size.
As the crystal size decreases, the measured optical band gap increases.
Some indirect semiconductors (such as silicon) may become illuminating, while some direct semiconductors
(such as gallium arsenide) may become indirect semiconductors.
These size effects of semiconductors' physical properties
pose significant challenges to conventional solid state physics:
The basic theory based on the Bloch theorem is incapable of explaining these size effects.
\par
A clear understanding of electronic states in a low-dimensional system or finite crystal
has theoretical and practical significance.
However, developing a general theory on electronic states in low-dimensional systems or
finite crystals with boundaries has always been considered difficult.
The main obstacle is that the atomic arrangement of a low-dimensional system
or a finite crystal does not have periodic translational invariance.
It is based on the atomic arrangement's periodic translational invariance that the Bloch theorem provides a theoretical
framework for the electronic states' theory in crystals in conventional solid state physics.
At the same time, it also dramatically simplifies the mathematics on solving relevant quantum mechanics equations.
Without such a theoretical framework and mathematical simplification, working on the electronic states
in low-dimensional systems or finite crystals with boundaries seems difficult.
Therefore, most previous theoretical investigations on electronic states in low-dimensional systems used
approximation methods or numerical approaches, usually for a specific material and a particular model.
One of the most frequently used approximation methods is the previously mentioned effective mass
approximation method.
\par
An ideal low-dimensional system or finite crystal is a simplified model of
a real low-dimensional system or finite crystal.
Based on the understanding of Bloch waves' quantum confinement, we have developed
an analytical theory of electronic states in ideal low-dimensional systems and
finite crystals.
By  “ideal”, it is assumed that:
(1) The potential
inside the low-dimensional system or finite crystal is the same as
in a crystal with periodic translational invariance;
(2) The electronic states are completely confined in
the limited size of the low-dimensional system or finite crystal.
These two simplifying assumptions facilitate the development of the analytical theory.
\par
This new theory abandons the necessary “periodic boundary condition(s)”
in conventional solid state physics.
As a single-electron and non-spin theory, this new theory can accommodate the
corresponding theory of electronic states in crystals in conventional solid-state physics,
thus can explain the problems that the traditional theory can explain.
Furthermore, it can also explain many problems that the conventional theory
cannot explain.
\par
The boundary-dependent ${\bf{\color{red} 1}}$ state and the $\textit{{\color{blue}(N~-~1)}}$
stationary Bloch states in Eq. (\ref{eqn:EQ1})
indicate that the electronic states in a low-dimensional system or a finite crystal
are not progressive Bloch waves as in conventional solid state physics.
They also distinguish this
new theory from conventional perceptions in the solid state physics community on electronic states
in low-dimensional systems or finite crystals (such as those from the effective mass approximation).
\par
On this basis, it is possible to explore and understand some of the most basic and
common physical problems of electronic states in low-dimensional systems and
finite crystals. The existence and property of surface states and other relevant
states are natural conclusions of this theory.
For example, in some most straightforward cases in a simple finite crystal or
quantum dot of a rectangular cuboid shape having $N_1$, $N_2$, $N_3$ periods
separately in three perpendicular directions,
for each bulk energy band, there are $(N_1-1)(N_2-1)(N_3-1)$ bulk-like
states, $(N_1-1)(N_2-1) + (N_2-1)(N_3-1) + (N_3-1)(N_1-1)$
surface-like states, $(N_1-1) + (N_2-1) + (N_3-1)$ edge-like states
and one vertex-like state; and the following general relations
exist:\par \hspace{10pt}The~energy~of~the~vertex-like~state~\par
\hspace{42pt}$>$~The~energy~of~every~edge-like~state~\par
\hspace{74pt}$>$~The~energy~of~every~relevant~surface-like~state~\par
\hspace{106pt}$>$~The~energy~of~every~relevant~bulk-like~state.
\par
Based on this theory, the earlier mentioned size effects and contradictions
between numerical calculations and effective mass approximation methods
could be understood and clarified.
This theory also offers new understandings of essential concepts in modern
solid state physics, such as surface states,  bandgap in the low-dimensional
system of cubic semiconductors, and others.
Besides, it also naturally raises new questions and makes new predictions.
\par
One of the predictions is that the fundamental difference between a macroscopic
semiconductor crystal and a macroscopic metal crystal may become blurred as the
crystal size becomes so small that the boundary-dependent states' effects become
significant to play a more substantial role. A low-dimensional system of a cubic
semiconductor may even have a metallic conductivity.
\par
Subsequent investigations by other authors have gradually confirmed some new conclusions
of the theory \cite{stmp2}.
\par
The mathematical theory has been extended to treat one-dimensional phononic crystals
and photonic crystals\cite{stmp2}. It is more straightforward and applicable than the
previously conventionally used transfer matrix methods. This approach provides new
theoretical understandings of relevant physics problems.

\par
Periodicity is one of the most fundamental and extensively investigated
mathematical concepts. Many periodic systems can be truncated. The truncations of
periodicity do raise new issues. Compared with the “periodicity”, the investigations
and general understandings of the “truncated periodicity” are much less.
Reported work is a preliminary attempt.
\par
This article is a brief introduction to the new theory. Readers who have further
interests in the new theory can refer to the author's original work \cite{syr1,stmp1,stmp2}.


\begin{thebibliography}{99}

\bibitem{lan}
For example, L. D. Landau and E. M. Lifshitz: \textit{Quantum Mechanics},
Pergamon Press Ltd, Oxford, (1958).

\bibitem{blh} F. Bloch, $\ddot{\rm U}$ber die quantenmechanik der elektronen in kristallgittern.  Zeit. Phys. {\bf 52}(7–8): 555–600 (1928).

\bibitem{eas} M. S. P. Eastham, {\it The Spectral Theory of
Periodic Differential Equations}, Scottish Academic Press,
Edinburgh/London, (1973).

\bibitem{syr1} Shang Yuan Ren, Two types of electronic states in
one-dimensional crystals of finite length,
Annals of Physics(NY) {\bf 301}: 22-30 (2002).

\bibitem{stmp1} Shang Yuan Ren, {\it Electronic States in
Crystals of Finite Size: Quantum confinement of Bloch waves}. First Edition, Springer, (2006).

\bibitem{stmp2} Shang Yuan Ren, {\it Electronic States in
Crystals of Finite Size: Quantum confinement of Bloch waves}. Second Edition. Springer, (2017).

\end{thebibliography}
\end{document}